\documentclass{article}

\usepackage{arxiv}

\usepackage[utf8]{inputenc} 
\usepackage[T1]{fontenc}    
\usepackage{hyperref}       
\usepackage{url}            
\usepackage{booktabs}       
\usepackage{amsfonts}       
\usepackage{nicefrac}       
\usepackage{microtype}      
\usepackage{booktabs}
\usepackage{tikz}
\usepackage[caption=false]{subfig}
\usepackage{graphicx}
\usepackage{amsmath}
\usepackage{url}

\DeclareMathOperator*{\argmax}{arg\,max}

\usetikzlibrary{positioning,shapes,calc,through,intersections,fit,matrix,patterns}

\tikzset{
  flowchart/.style={
      process/.style={rounded rectangle, draw,align=center},
      decision/.style={diamond, aspect=2, draw,align=center},
      transition/.style={draw,->},
      dataflow/.style={->,dashed},
    },
    mnemonic/.style={
      label={[fill=blue!30]120:#1}
    },
}

\title{A Generic Metaheuristic Approach to Sequential Security Games}

\author{
  Adam {\.Z}ychowski \\
  Faculty of Mathematics and Information Science, \\
  Warsaw University of Technology \\
  Warsaw, Poland\\
  \texttt{a.zychowski@mini.pw.edu.pl} \\
   \And
  Jacek Ma{\'n}dziuk \\
  Faculty of Mathematics and Information Science, \\
  Warsaw University of Technology \\
  Warsaw, Poland\\
  \texttt{j.mandziuk@mini.pw.edu.pl} \\
}

\begin{document}
\maketitle

\begin{abstract}  
The paper introduces a generic approach to solving Sequential Security Games (SGs) which utilizes Evolutionary Algorithms. Formulation of the method (named \emph{EASG}) is general and largely game-independent, which allows for its application to a wide range of SGs with just little adjustments addressing game specificity. Comprehensive experiments performed on $3$ different types of games (with $300$ instances in total) demonstrate robustness and stability of EASG, manifested by repeatable achieving optimal or near-optimal solutions in the vast majority of the cases. The main advantage of EASG is time efficiency. The method scales visibly better than state-of-the-art approaches and consequently can be applied to SG instances which are beyond capabilities of the existing methods. Furthermore, due to \emph{anytime} characteristics, EASG is very well suited for time-critical applications, as the method can be terminated at any moment and still provide a reasonably good solution - the best one found so far.
\end{abstract}


\section{Introduction}
\label{sec:introduction}

Game theory (GT) offers mathematically justified solutions to many practical problems arising in various domains, e.g. economics~\cite{friedman1986game}, biology~\cite{parker1990optimality}, politics~\cite{mccarty2007political} or social networks~\cite{slikker2012social}. One of the fast-growing GT areas is security management (e.g. border controlling, police surveillance, anti-terrorism policy, patrolling schedules, etc.). The majority of these applications base on the concept of Security Games.

Security Games (SGs) are typically two-player Stackelberg Games \cite{von1934marktform}. One of the players - the \emph{leader} (a.k.a. the \emph{Defender} in SGs) commits to a strategy first and then the other player - the \emph{follower} (a.k.a. the \emph{Attacker} in SGs) chooses their strategy - based on the already announced leader's strategy. Such a sequence of decisions introduces asymmetry of information access, favoring the follower. The above concept fits certain real-life security-related scenarios in which the Attacker can surveil the guards (playing the Defender's role) patrolling some critical area and discover / deduct their patrolling strategy.

The list of successfully deployed real-world applications of SGs includes the scheduling system for Los Angeles International Airport canine patrols~\cite{jain2010software}, the PROTECT system which randomizes schedules of US Coast Guard's resources in Boston harbour~\cite{shieh2012protect}, the TRUSTS system for scheduling patrols for fare inspection in Los Angeles Metro system~\cite{yin2012trusts}, or the PAWS system to prevent poaching and protecting wildlife in Queen Elizabeth National Park in Uganda~\cite{yang2014adaptive}. Please consult a recent survey paper~\cite{sinha2018stackelberg} for other examples.

In SGs the goal is to find a pair of players' strategies that correspond to the \emph{Stackelberg Equilibrium} (StE)~\cite{leitmann1978generalized}. Since the Attacker is aware of the opponent's strategy in advance, their strategy is a consequence of the Defender's strategy (they choose a strategy which provides them the highest possible payoff). Hence, the problem of finding StE in SGs can be reduced to finding an optimal Defender's strategy. While the task is trivial in the case of pure strategies, in the mixed strategies space it becomes NP-hard~\cite{conitzer2006computing}.

On a general note, there are two major types of SGs solution methods: exact and approximate. Majority of the proposed exact approaches employ Mixed-Integer Linear Programming (MILP)~\cite{kiekintveld2009computing, jain2010software, bosansky2015sequence, cermak2016using} and, consequently, suffer from poor time scalability which hinders their application beyond a certain level of game's complexity. Approximate methods, on the other hand, offer much better scalability but at the cost of finding close-to-optimal solutions~\cite{kumar2010existence, kiekintveld2012efficient, vcerny2018incremental, karwowski2019stackelberg, karwowski2019monte, wang2019deep}. This paper follows the latter approach and adopts Evolutionary Algorithms metaheuristic~\cite{back1997handbook}, which has proven efficient in solving other types of bi-level optimization problems~\cite{colson2007overview,bard2013practical}.

Evolutionary Algorithms (EAs) are inspired by the process of (biological) evolution. Each member of a population in EA represents a candidate (valid) solution. In the iterative process of creating new generations three evolution-related operations: mutation, crossover and selection are performed. Despite the lack of rigorous proofs of convergence to the optimal solution, EAs - thanks to their experimentally proven efficiency - have been widely applied to a wide range of real-life optimization problems, e.g.~\cite{coello2004applications, hu2015application, mandziuk2016memetic, zychowski2018addressing}.

In the context of GT, Sefrioui et al.~\cite{sefrioui2000nash} applied co-evolutionary algorithm for computing the Nash equilibrium on a fluid dynamics problem and compared the results with Pareto Genetic Algorithms~\cite{rey1994niched}. A similar approach, relying on colonial competitive algorithm~\cite{atashpaz2007imperialist}, was proposed in~\cite{rajabioun2008colonial}. There were also certain attempts of using EAs for finding StE~\cite{vallee1999off, sakawa2000computational, camacho2014solving}, however, all of them were designed for simple one-step games. To our knowledge, the only successful application of evolutionary methods to sequential SGs was proposed in~\cite{karwowski2019memetic}, however this approach is domain-dependent and its application is limited
to a specific genre of SGs, i.e. games played on a plane with moving targets.

\subsection{Contribution}
\label{sec:Contribution}
The main contribution of this paper is an introduction of the first evolutionary approach to finding near-optimal solutions of a wide range of SGs in a computationally efficient manner. The method, abbreviated as EASG (Evolutionary Algorithm for Security Games), is domain-independent and flexible, i.e. applicable to various SGs formulations (one-step, multi-step, general-sum, with limited observability, with bounded rationality, etc.) with little adjustments only. Furthermore, due to \emph{anytime} characteristics, EASG is specifically well suited for time-critical applications in SG domain.

EASG is tested on three types of SGs, where it demonstrates optimal or close-to-optimal performance, while excelling the competitive approaches in terms of time scalability.

The remainder of this paper is arranged as follows. Section~\ref{sec:problem} presents a formal description of sequential Security Games. Section~\ref{sec:state-of-the-art} provides an overview of the state-of-the-art approaches to SGs, used in the experimental evaluation of EASG. A detailed description of all EASG components is presented in the next section. Section~\ref{sec:setup} is devoted to experimental setup (EASG parametrization and tuning, as well as description of benchmark games), which is followed by presentation of experimental results in Section~\ref{sec:evaluation}. The last section contains conclusions.


\section{Problem definition}
\label{sec:problem}

We consider $n$ step games with two players: the Defender ($D$) and the Attacker ($A$). In each time step ($\tau_1,\ldots,\tau_n$) player $p$ chooses action $a^p_i$ ($p \in \{D,A\}, i = 1, \ldots, n$) from the set of available actions $M(s^p_i)$ where $s^p_i$ is a state of player $p$ in time step $\tau_i$. State $s^p_i$ is determined by previous player's actions, his initial position and the opponent's actions. Players are not aware of the opponent's actions.

For each game state $s$ there are four predefined payoffs $U^k(s)$ $(k \in \{A+, A-, D+, D-\})$ representing the Attacker's reward ($U^{A+}(s)$), their penalty ($U^{A-}(s)$), the Defender's reward ($U^{D+}(s)$) and their penalty ($U^{D-}(s)$), resp. Some of the states (usually those with high $U^{A+}(s)$ values) are distinguished and called the targets.

If in any time step $\tau_i (i=1,\ldots,n)$:
\begin{itemize}
\item The Attacker and the Defender move to the same state (say $s_i$), then the game ends (the Attacker is intercepted) and the players receive payoffs $U^{A-}(s_i)$ and $U^{D+}(s_i)$, resp.
\item  The Attacker reaches any of the targets (say $s_j$) and is not intercepted, then the game ends and the respective payoffs are equal to $U^{A+}(s_j)$ and $U^{D-}(s_j)$.
\end{itemize}
Otherwise, the game lasts for $n$ time steps and ends with neutral payoffs.

A \emph{pure strategy} of the player is defined as an assignment of one action to each \emph{potentially reachable} state of the game. Please observe that the same action of the player may lead to several reachable states depending on the opponent's action selection~\cite{kuhn1950extensive}.

Let's denote a set of all pure strategies of player $p$ by $\Sigma^p$. A \emph{mixed strategy} $\pi^p \in \Pi^p$, where $\Pi^p$ is a set of all possible mixed strategies of player $p$, is a probability distribution over $\Sigma^p$.

The game model employs Stackelberg Game (StG) principles which means that first the Defender commits to their strategy $\pi^D \in \Pi^D$ (probability distribution of their moves) and then the Attacker, being aware of $\pi^D$, determines their strategy $\pi^A$.

Let's denote by $U^p(\pi^D,\pi^A)$ an expected utility value of player $p$ as a result of the game played according to mixed strategies $\pi^D$ and $\pi^A$. Stackelberg Equilibrium can be formally defined as a pair ($\pi^D$, $\pi^A$) satisfying the following equations:
\begin{equation}
BR(\pi^D) = \argmax_{\pi^A \in \Pi^A} U^A(\pi^D, \pi^A)
\label{eq:SE1}
\end{equation}
\begin{equation}
\argmax_{\pi^D \in \Pi^D} U^A(\pi^D, BR(\pi^D))
\label{eq:SE2}
\end{equation}
Equation (\ref{eq:SE1}) defines the Attacker's best (optimal) response $BR(\pi^D)$ to the Defender's strategy $\pi^D$ while eq. (\ref{eq:SE2}) selects the best Defender's strategy against the optimal Attacker's response. In order to avoid unambiguity, if there exist more than one best Attacker's response, Strong Stackelberg Equilibrium (SStE) was defined~\cite{leitmann1978generalized} in which the Attacker, among all their best responses (with the same highest utility for them), selects the one with the highest Defender's utility, i.e. breaks ties in favor of the Defender. In this paper SStE version of StE is considered.


\section{State-of-the-art Approaches}
\label{sec:state-of-the-art}

The vast majority of hitherto approaches to solving StG were focused on one-step game formulations and either couldn't be applied to the multi-step case considered in this paper, or their application would be highly inefficient. In the literature there is just a handful of methods devoted to multi-step extensive-form StG, all of them developed in the last few years.
%
%

The first notable approach (BC2015) was introduced by Bo\u sansk\'y and \u Cerm\'ak in 2015~\cite{bosansky2015sequence}. The authors extended a very popular MILP-based method DOBBS~\cite{paruchuri2008playing} to extensive-form games and introduced a novel algorithm for computing SStE, able to exploit the underlying structure of sequence-form games. The method is designed for non-zero-sum games and reduces the size of a linear program by transforming an extensive-form game into its equivalent sequence-form representation. In effect, the the size of a linear program is reduced from exponential (as in DOBBS) to linear with respect to the game tree size, but still exponential with respect to the game length. An exact formulation of MILP utilized in BC2015 can be found in~\cite{bosansky2015sequence}.

Another approach to exact computation of SStE for two-player extensive-form general-sum games was proposed by \u Cerm\'ak et al.~\cite{cermak2016using}. The method (henceforth referred to as C2016) utilizes a correlated version of SStE known as Stackelberg Extensive-Form Correlated Equilibrium (SEFCE). In SEFCE, the Defender can send signals to the Attacker who must follow these signals in their choice of the best response. C2016 relies on a linear program for computing SEFCE and then modifies it by iteratively restricting the signals the Defender can send to the Attacker and thus converging to SStE. In the experimental evaluation presented in~\cite{cermak2016using}, C2016 was superior to BC2015 in terms of time efficiency.

One of the recent heuristic approaches to SStE approximation was the Mixed-UCT method~\cite{karwowski2015new, karwowski2019monte} which incorporates a variant of Monte Carlo Tree Search, known as Upper Confidence Bounds~\cite{kocsis2006bandit}. The authors~\cite{karwowski2019monte} subsequently proposed another UCT-based approach - O2UCT~\cite{karwowski2019stackelberg,karwowski2020AAAI} which relies on a guided sampling of the Attacker's strategy space interleaved with finding (using double-oracle method~\cite{jain2011double}) a feasible Defender's strategy for which the just-sampled Attacker's strategy is the optimal response. Experimental evaluation of O2UCT shows the method's ability to find optimal or close-to-optimal solutions for various types of test games and better time-scalability than both the above mentioned exact MILP methods.

Another approximate method (denoted by CBK2018)~\cite{vcerny2018incremental} is a heuristic time-optimized MILP approach which constructs a smaller game tree representation with a specific abstraction structure called \emph{gadgets}. The method significantly reduces the Defender's strategy space and therefore, brings down computation time requirements, however at the cost of loosing theoretical MILP property of convergence to the optimal solution. A diminished game is solved with the C2016 method.

Four algorithms summarized above (two exact: BC2015, C2016 and two approximate:O2UCT, CBK2018) are state-of-the-art approaches to solving multi-step extensive-form StG and will be used as reference methods in the experimental evaluation of EASG.


\section{Evolutionary Algorithm for Security Games}
\label{sec:EASG}
This section contains a detailed description of the proposed Evolutionary Algorithm for Security Games (EASG).

\subsection{System overview}
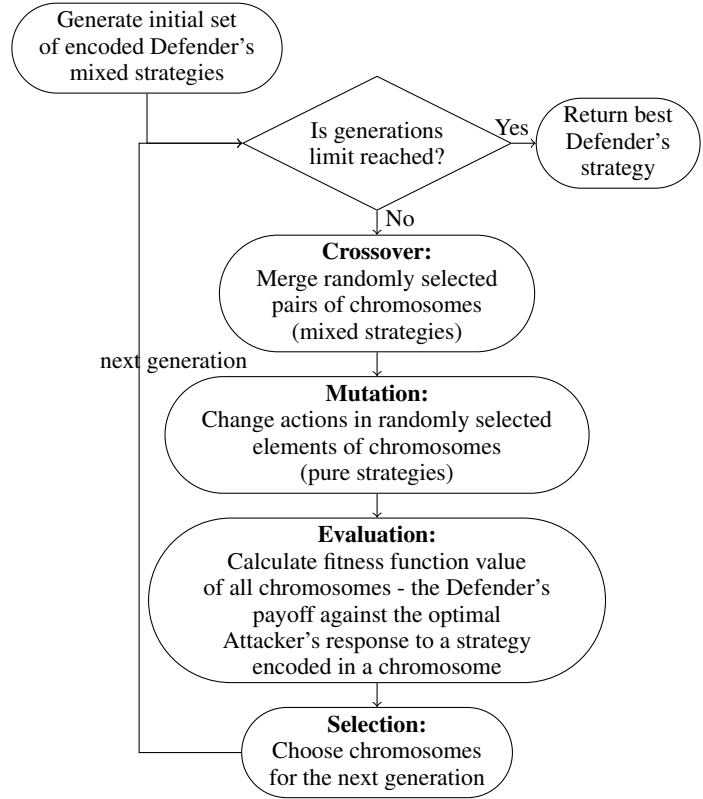
\begin{figure}[ht]
\begin{center}
\begin{tikzpicture}[
  flowchart,
  dsmutate/.style={fill=red!30},
  node distance=1.0em,
  description/.style={draw, dashed, align=center, inner sep=0.01}
  ]
  \small
  \node(init)[process] {Generate initial set \\of encoded Defender's \\mixed strategies};

  \node[decision,right=of init, yshift=-4em, xshift=-2.7em] (stop_condition) {Is generations \\limit reached?};

  \node(crossover)[process,below=of stop_condition] {\textbf{Crossover:} \\ Merge randomly selected\\ pairs of chromosomes \\ (mixed strategies)};
  \node(mutation)[process,below=of crossover] {\textbf{Mutation:}\\Change actions in randomly selected \\elements of chromosomes \\(pure strategies)};
  \node(evaluation)[process,below=of mutation] {\textbf{Evaluation:}\\Calculate fitness function value\\ of all chromosomes - the Defender's \\ payoff against the optimal \\ Attacker's response to a strategy \\ encoded in a chromosome};
  \node(selection)[process,below=of evaluation] {\textbf{Selection:}\\Choose chromosomes \\for the next generation};
  \node(return)[process,right=of stop_condition] {Return best \\ Defender's\\ strategy};

  \draw[transition] (init) |- (stop_condition);
  \draw[transition] (stop_condition) -- (crossover)node[pos=0.3,auto]{No};
  \draw[transition] (crossover) -- (mutation);
  \draw[transition] (mutation) -- (evaluation);
  \draw[transition] (evaluation) -- (selection);
  \draw[transition] (selection) -- ++(-10em,0) |-
  (stop_condition)node[pos=0.32,auto, xshift=4.9em]{next generation};
  \draw[transition] (stop_condition) -- (return)node[pos=0,auto]{Yes};



\end{tikzpicture}
\caption{An overview of EASG.}
\label{fig:overview}
\end{center}
\end{figure}

Figure~\ref{fig:overview} presents a flow-chart of the algorithm. Initially, a population of $p_{size}$ individuals (chromosomes) is randomly generated. Each individual represents a potential solution - Defender's mixed strategy. In each subsequent generation, crossover and mutation operators are applied to randomly selected chromosomes from the current population, which is then followed by a selection procedure that promotes individuals to the next generation based on their evaluation (fitness). The fitness of a given chromosome is calculated as the Defender's payoff (when the Defender plays a mixed strategy represented by that chromosome) against an optimal Attacker's response (an Attacker's pure strategy yielding the highest payoff for them).

The above procedure is executed until the best (found so far) Defender's payoff does not change within $n_c$ iterations or the limit for the number of generations $n_g$ is exceeded.

\subsection{Chromosome representation}

Each chromosome (individual) represents some Defender's mixed strategy (a candidate SStE solution) in the form of a vector of pure strategies $\pi^q_i$ and their respective probabilities $p^q_i$:
\begin{equation}CH_q = \{(\pi^q_1,p^q_1), \ldots, (\pi^q_{l_q},p^q_{l_q})\},\ \ \sum_{i=1}^{l_q}p_{l_q}^q=1,
\label{eq:chromosome}
\end{equation}
where $l_q$ is the length of chromosome $CH_q$, i.e. the number of pure strategies included in the mixed strategy represented by that chromosome. A particular form of a pure strategy dependents on game specificity. In the most common case, a pure strategy is represented as a list of Defender's actions in consecutive time steps.

\subsection{Initial population}

Initial population contains solely pure strategies, i.e. $\forall_q\  l_q = 1\  \land\  p^q_1 = 1$. These strategies are generated randomly in the following way. For each strategy, in each time step the next action is selected uniformly from all actions available in a given state. This procedure is independently executed for each chromosome in the initial population, i.e. $p_{size}$ times.

\subsection{Crossover}

In the first step of crossover operation, a subset of $p_c\cdot p_{size}$ individuals are randomly selected from the population, where $p_c$ is crossover rate. Then, individuals from this subset are randomly paired (in the case of an odd number of individuals, a randomly chosen one is removed). From each pair of individuals one new \emph{offspring} chromosome is created in the following way. All pure strategies from the \emph{parent} chromosomes are merged into one mixed strategy with their probabilities halved. Next, each pure strategy $\pi^q_i$ in this newly created chromosome, except for the one with the highest probability, is removed with probability $(1-p^q_i)^2$ (i.e. the lower the probability of a strategy the higher its chance for being deleted). Afterwards, probabilities of the remaining pure strategies are normalized so as to sum up to $1$. Note that the \emph{offspring} chromosome inherits all pure strategies from both its parents and without such a reduction, evaluation of the chromosome (computation of the respective Defender's payoff) would be inefficient, due to a high number of pure strategies with low probabilities.

\begin{figure}[ht]
	\begin{center}
    \includegraphics[width=0.5\columnwidth]{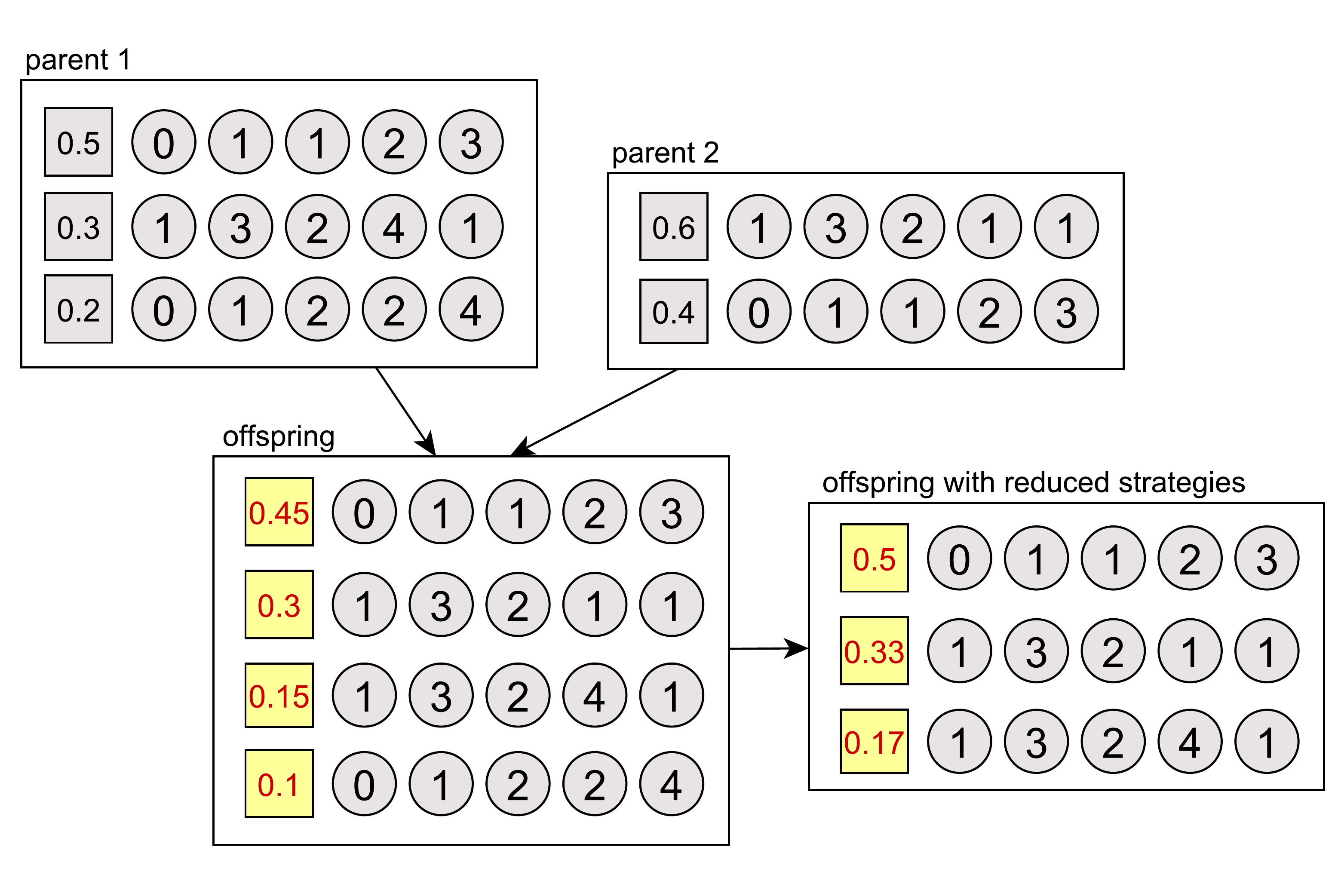}
    \caption{An example of crossover operation. Each row of circles represents a sequence of actions in consecutive time steps (a Defender's pure strategy), squares denote probabilities assigned the respective strategies. In the first step an offspring chromosome contains all pure strategies inherited from the parents (with halved probabilities). Then some strategies are potentially removed (in the example it is the one with the lowest probability) and probabilities of the remaining strategies are normalized (to sum up to $1$).}
    \label{fig:crossover}
	\end{center}
\end{figure}

The role of the crossover operator is to enhance exploitation aspect of the EA by means of mixing strategies found so far. A sample crossover operation is presented in Fig.~\ref{fig:crossover}.

\subsection{Mutation}

The mutation operator is applied to each chromosome independently with probability $p_m$. First, one pure strategy in the chromosome is randomly chosen. Then iteratively, starting from a randomly selected time step $t_i$ up to the last time step $t_n$, an action in a considered time step $t_j, i\le j\le n$ is changed to an action uniformly chosen among all actions available in the corresponding game state.

\begin{figure}[ht]
	\begin{center}
    \includegraphics[width=0.5\columnwidth]{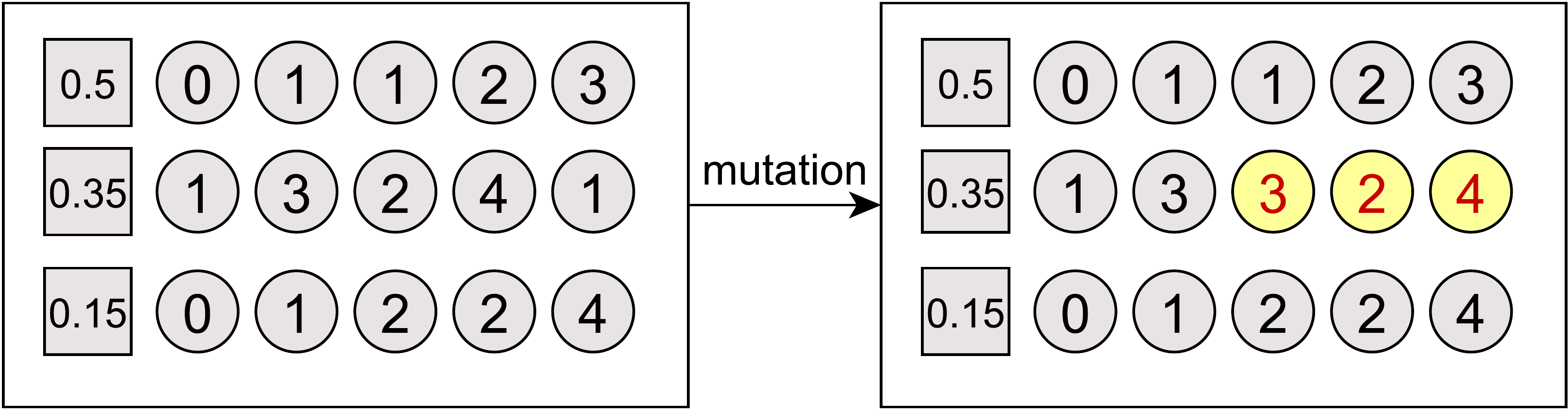}
    \caption{An example of mutation operation. Each row of circles represents a sequence of actions in consecutive time steps (a Defender's pure strategy), squares denote probabilities assigned the respective strategies. In the presented case, the second pure strategy is affected by the mutation operation, starting from the third time step.}
    \label{fig:mutation}
	\end{center}
\end{figure}

The role of the mutation operator is to boost exploration of the new areas of the search space. An example of mutation operation is presented in Fig.~\ref{fig:mutation}.

\subsection{Evaluation}

The fitness of a chromosome is calculated as a Defender's payoff when they play a mixed strategy encoded in that chromosome. Since it is proven~\cite{conitzer2006computing} that in StG there always exists at least one pure strategy of the follower which is their best response to the leader's strategy (pure or mixed), then it is sufficient to iteratively check all pure Attacker's strategies and select the one with the highest Attacker's payoff (additionally breaking ties in favor of the Defender - SStE condition). The Defender's payoff against the best Attacker's response found in the above described way is then computed and returned as a chromosome fitness value.

\subsection{Selection}

In the first step of the selection procedure, $e$ individuals from the current population with the highest fitness values (including those created by mutation or crossover operations) are unconditionally promoted to the next generation population. These individuals are called \emph{elite} and transmit the top ranked solutions found so far through consecutive generations.

Next, a \emph{binary tournament} is iteratively executed until the next generation population is filled with $p_{size}$ individuals. In each iteration, among all individuals in the current generation (including the mutated or crossed-over individuals) two chromosomes are randomly chosen and, from this pair, the one with a higher fitness value is promoted to the next generation with probability $p_s$, otherwise the lower-fitted one is promoted.


\section{Experimental setup}
\label{sec:setup}

\subsection{Benchmark games}

Properties of EASG were tested on $3$ sets of multi-step games with variable characteristic: Warehouse Games~\cite{karwowski2019monte}, Search Games~\cite{bosansky2015sequence}, and FlipIt Games~\cite{van2013flipit}, which were previously used in the literature for evaluation of various state-of-the-art SG approaches. While all $3$ types of games are defined on graphs, their rules, properties and related challenges are vastly different, which makes them a truly diverse benchmark set.

\subsubsection{Warehouse Games}
The Warehouse Games (WHG) proposed in~\cite{karwowski2019monte} are played on graphs which mimic the layouts of warehouse buildings. Vertices represent storage rooms and corridors. Attacker and Defender start in fixed (different) vertices. In each time step, each player can move to any of the neighboring vertices (directly connected through an edge) or stay in the currently occupied vertex. The game ends in any of the following three cases:
\begin{description}
\item[a)] players meet in the same vertex, which means interception of the Attacker - a negative payoff for the Attacker and a positive one for the Defender,
\item[b)] the Attacker reaches one of the distinguished vertices (targets) and is not intercepted - the Attacker receives a positive payoff and payoff of the Defender is negative,
\item[c)] none of the two above cases takes place in any of the $n$ time steps - the payoff equals $0$ for both playing sides.
\end{description}
The payoff structure is non-zero-sum. A more detailed description of these benchmark games is presented in~\cite{karwowski2019monte}. The number of steps in WHG varies between $3$ and $8$ which leads to the game trees of the sizes from $10^2$ to $10^8$ nodes. For each number of steps $t=3,\ldots,8$, $25$ games were tested ($150$ different games in total). All test games were downloaded from the website~\cite{sg-mini}.

In WHG an action is in the form of a decision about the next node to be visited, hence in EASG application to WHG a chromosome contains a list of nodes to be visited in consecutive time steps.

\subsubsection{Search Games}
The Search Games (SEG)~\cite{bosansky2015sequence} are played on directed graphs. The Attacker's goal is to reach one of the distinguished target vertices, starting from a fixed initial vertex. Contrary to WHG, in SEG Defender controls several units and furthermore these units cannot move freely on the entire graph but each of them has a subset of vertices assigned which it is allowed to visit.

Another crucial difference compared to WHG is the property of partial observability. Namely, the Attacker leaves traces in visited vertices which can be discovered by a Defender's unit if they visit the node after the Attacker's presence (in one of the subsequent time steps). However, the Attacker has the ability to erase such a trace if they spend an additional round in a given vertex (i.e. stay in this vertex in two or more consecutive time steps).

The end-of-game conditions are the same as in the WHG - the Defender obtains a positive payoff for catching the Attacker, or the Attacker is rewarded for reaching a target vertex without being intercepted, or the game ends with neutral payoffs after a certain number of rounds.

In total, $90$ games with $4$, $5$, and $6$ time steps, played on $3$ different graph structures proposed in~\cite{bosansky2015sequence} were used in the evaluation process. Attacker's and Defender's penalties were set to $-1$ and for each graph structure $10$ random distributions of rewards were sampled from interval $[1,2]$.

In order to be able to solve SEG instances the EASG chromosome representation needs to be extended.
Formally, let's denote by $\mathcal{L}^u_{vt}$ a list of nodes visited in consecutive time steps by unit $u$ in the case of trace discovery in node $v$ at time step $t$ and by $\mathcal{L}^u_{\emptyset}$ a list of nodes visited by unit $u$ in the case of no trace discovery during the whole gameplay.
Then, the Defender's pure strategy $\pi^D$ in SEG is of the following form: $\pi^D = \{ \mathcal{L}^u_{vt}, u \in D_u, v \in V, t \in \{1,\cdots,n\} \} \cup \{\mathcal{L}^u_{\emptyset}, u \in D_u\}$, where $D_u$ is a set of Defender's units, $V$ is a set of graph vertices, $n$ is the number of game steps.

Each pure strategy represents one of the possible compound scenarios, i.e. refers to all Defender's units. For each unit such a scenario is related to discovering or not (by that unit) a trace in any particular vertex, in any particular time step during the entire game. A set of such pure strategies with assigned probabilities defines a mixed strategy for the Defender, represented by a chromosome (cf. eq.~(\ref{eq:chromosome})).

\subsubsection{FlipIt Games}
The FlipIt Games (FIG)~\cite{van2013flipit} were initially proposed for evaluation of CBK2018 method~\cite{vcerny2018incremental}. FIG instances refer to cybersecurity settings in which the Attacker attempts to gain control over (infect) certain resources (e.g. servers, hubs, individual computers), and the Defender may take actions to restore their control on these infected units. The game is played on a directed graph for a fixed number of rounds.

In each round, each of the players selects a node he/she attempts to take control of (to \emph{flip the node}). Only some of the nodes (so-called entry nodes) are publicly accessible and the Attacker is obliged to start gaining control and penetrate the network from one of these nodes (in other words the only possible Attacker's action in the first round is selection of one of the entry nodes). A flip action is successful if the player controls also at least one of the preceding nodes (or it is the entry node with no predecessors) and the current owner of the selected node does not take the flip action in this node in the same round. Successful flip action results in gaining control on the flipped node.

At the beginning of the game all nodes are controlled by the Defender. Each node has assigned some cost, which is to be paid at each flip attempt, and a reward for controlling it. The final player's payoff is the sum of rewards from all nodes controlled by him/her after each round (nodes controlled in multiple rounds count multiple times), decreased by the costs of all flip attempts (either successful or failed).

In the experiments, $60$ FIG instances played on $3$ different graph structures proposed in~\cite{vcerny2018incremental} were used. For each graph, $5$ different payoffs structures were randomly drawn. The number of rounds was set to $3, 4, 5$ or $6$. The experiments were performed in \emph{No-Info} game variant~\cite{vcerny2018incremental} in which players have no information about the results of their actions - they are not aware of whether or not their flip action succeed and, therefore, their strategy is independent of the opponent's actions.

FIG are solved with the same EASG settings as in the case of WHG, with no specific adaptation. In particular, the form of a chromosome does not change, i.e. strategies are represented as lists of nodes which the Defender attempts to flip in consecutive game rounds.

\subsection{EASG parametrization}
\label{sec:parameters_tuning}

\begin{figure*}
\centering
  \subfloat[Population size ($p_{size}$).]{
    \includegraphics[width=.35\columnwidth]{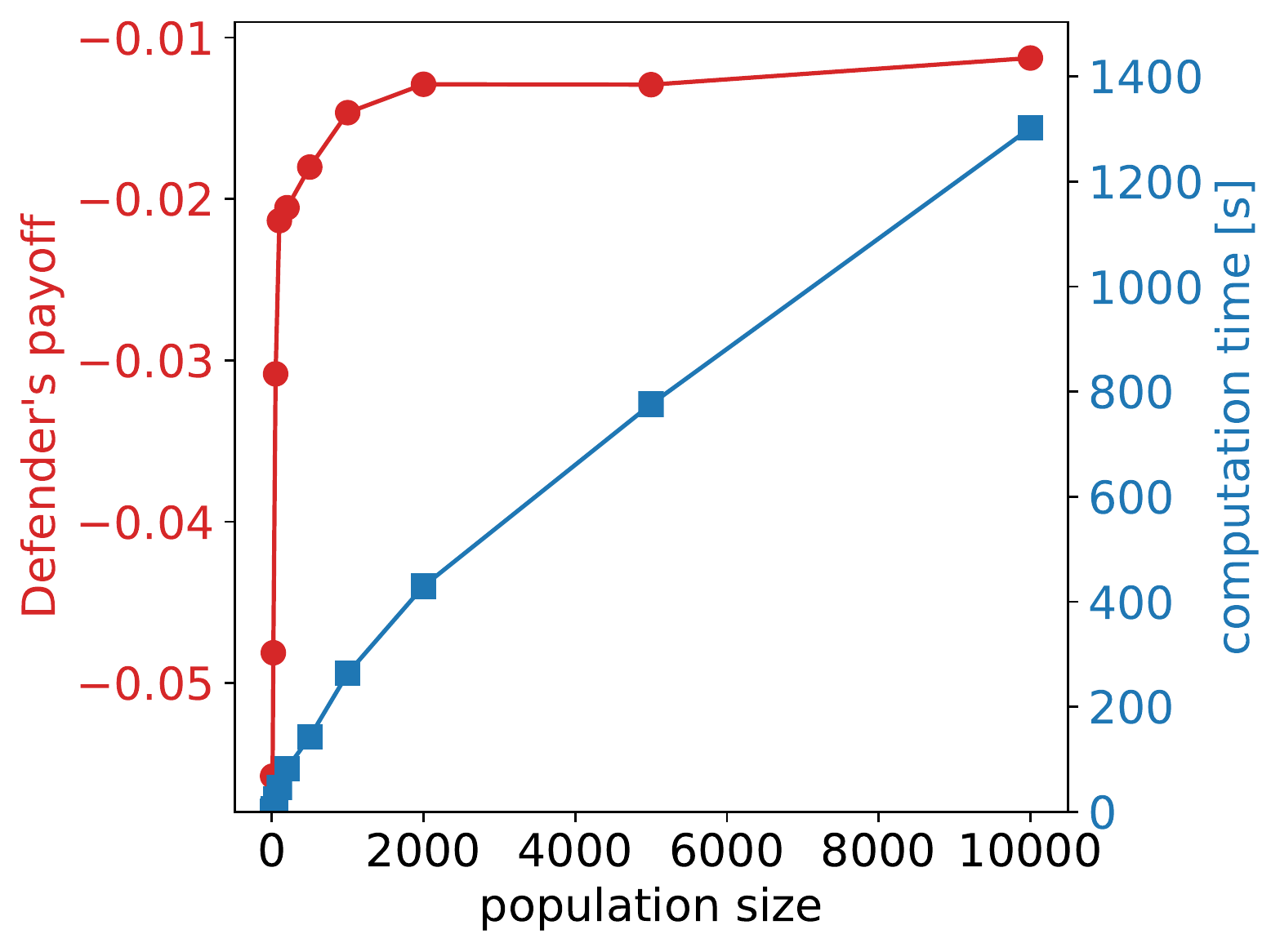}%
    \label{fig:population_size}
  }
  \quad
  \subfloat[Mutation rate ($p_m$).]{
    \includegraphics[width=.35\columnwidth]{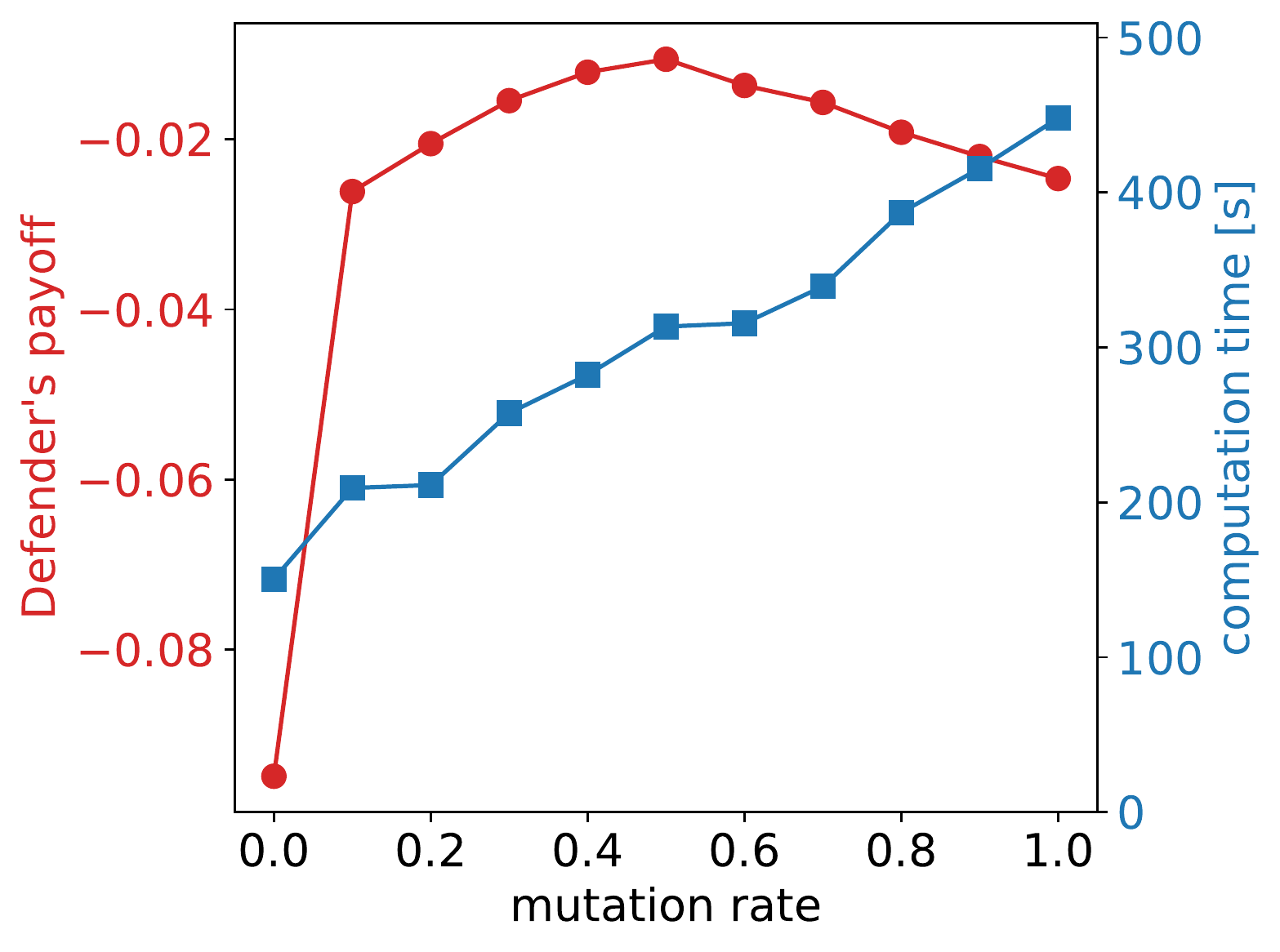}%
    \label{fig:mutation_rate}
  }
  \quad
  \subfloat[Crossover rate ($p_c$).]{
    \includegraphics[width=.35\columnwidth]{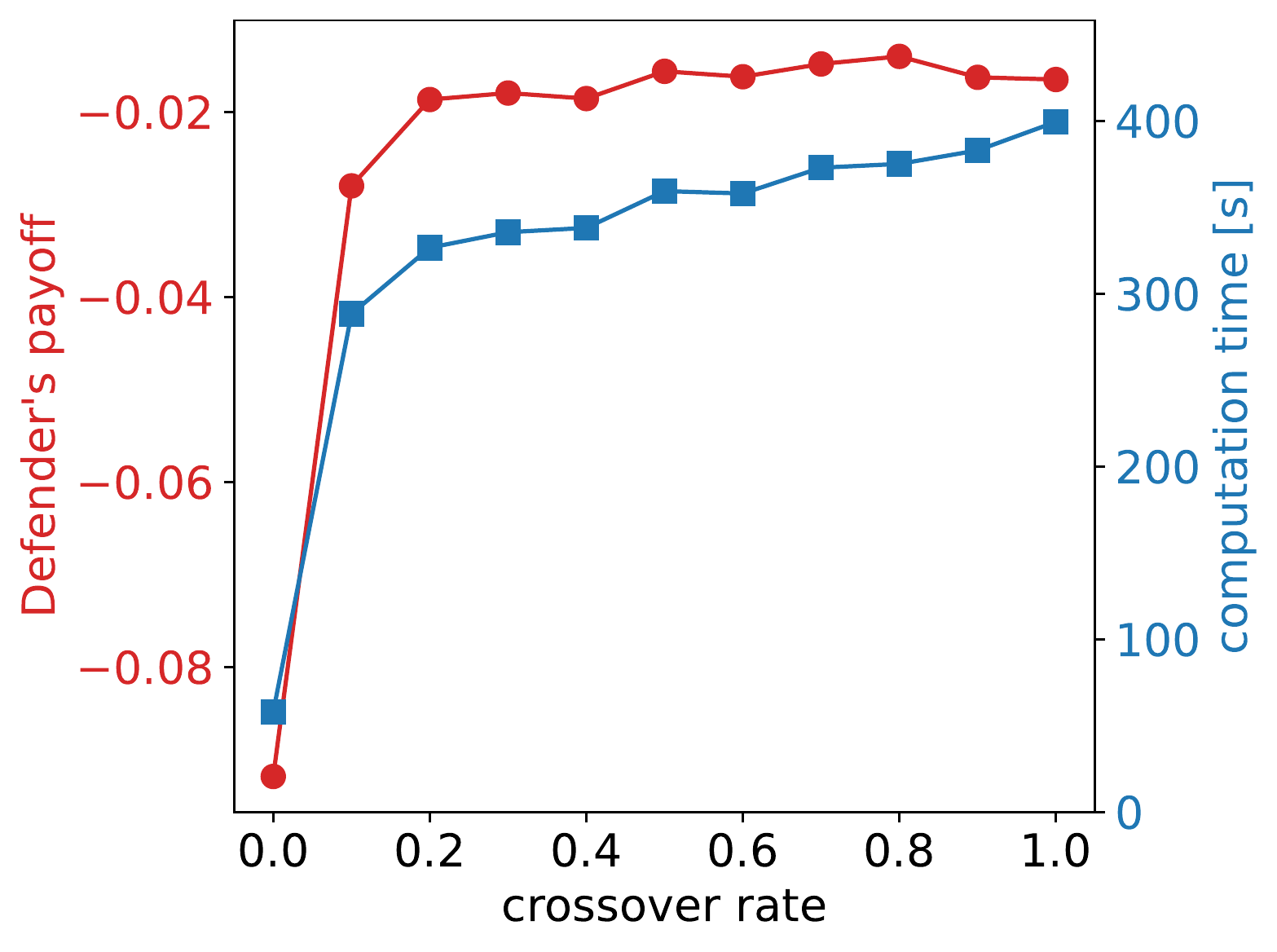}%
    \label{fig:crossover_rate}
  }
  \quad
  \subfloat[Selection pressure ($p_s$).]{
    \includegraphics[width=.35\columnwidth]{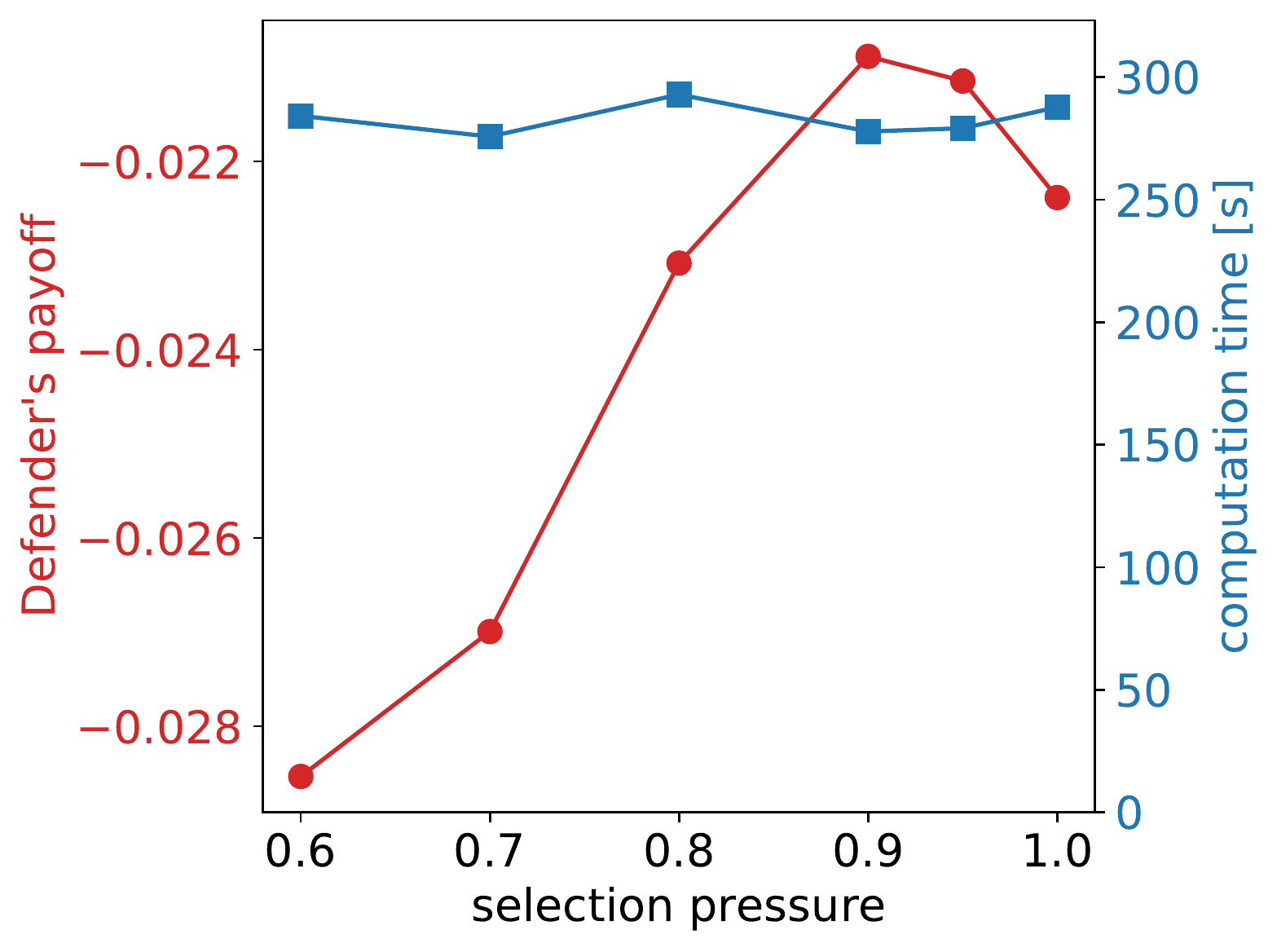}%
    \label{fig:selection_rate}
  }
  \caption{The average Defender's payoff (circles) and computation time (squares) with respect to the main steering parameters of EASG, calculated during the parameter tuning phase.}%
  \label{fig:all-params}
\end{figure*}

EASG parameters were tuned on $50$ WHG with $5$ and $6$ steps (excluded from the final tests). The initially selected sets of parameter values are presented in Table~\ref{tab:parameters}. EASG was run $5\;000$ times, in each case with a random selection of parameters (from Table~\ref{tab:parameters}) and a randomly chosen game.
Figures.~\ref{fig:population_size}-~\ref{fig:selection_rate} present Defender's payoff and computation time for the respective tested parameter averaged across all trails. Based on these outcomes the final parameter values (bolded in Table~\ref{tab:parameters}) were chosen, taking into account both the expected payoff and the running time.
%
\begin{table}[h]
\small
\centering
  \begin{tabular}{l|c|l}
    parameter&symbol&value\\
    \hline
    population size & $p_{size}$&$10, 20, 50, \textbf{100}, 200, 500, 1000, 2000, 5000$\\
    \# generations&$n_g$&$\textbf{1000}$\\
    \# generations with& & \\
    no improvement&$n_c$&$\textbf{20}$\\
    mutation rate & $p_{m}$&$0, 0.1, 0.2, 0.3, 0.4, \textbf{0.5}, 0.6, 0.7, 0.8, 0.9, 1$\\
    crossover rate & $p_{c}$&$0, 0.1, 0.2, 0.3, 0.4, 0.5, 0.6, 0.7, \textbf{0.8}, 0.9, 1$\\
    selection pressure & $p_{s}$&$0.6, 0.7, 0.8, \textbf{0.9}, 0.95, 1$\\
    \# elite & $e$ & $\textbf{2}$\\
  \end{tabular}

\caption{Parameter values selected for the tuning process. Finally recommended values are bolded.}%
\label{tab:parameters}
\end{table}

Figure~\ref{fig:population_size} shows that increase of \emph{population size} results in higher Defender's payoffs, which is quite intuitive as more individuals can explore strategy space and consequently the higher number of potential solutions is considered. For small values ($10\le p_{size}\le 100$) a steep increase of the expected payoff can be observed, which subsequently flattens. Computation time scales approximately linearly with $p_{size}$.

%
%

A relation between the Defender's payoff and \emph{mutation rate} is presented in Fig.~\ref{fig:mutation_rate}. Mutation is a key element of the EA process and without it ($p_m=0$) the Defender's payoff deteriorates drastically. On the other hand, too high mutation rate also entails payoff decrease, since the process is too disordered - the majority of strategies are randomly changed in each generation. Computation time naturally grows along with the mutation rate increase due to increasing number of chromosome modifications, but the increase is not as significant as in the case of population size (Fig.~\ref{fig:population_size}).
%

Similarly to mutation, \emph{crossover} is recognized as a critical operator in the EA process and its removal ($p_c=0$) causes significant payoff decrease (Fig.~\ref{fig:crossover_rate}). At the same time, differences among payoffs for all other (positive) crossover rate values are not significant. Clearly, the higher the $p_c$ the more crossover operations, and consequently the higher computation time, but the time increase is not steep.
%

The \emph{selection pressure} $p_s < 0.5$ would mean that weaker individuals were more likely to be chosen. Hence, only $p_s > 0.5$ were tested with the best results obtained for $p_s=0.9$ (see Fig.~\ref{fig:selection_rate}). Since this parameter does not affect the number of EASG operations, computation time remains approximately constant for all values of~$p_s$.
%
%


\section{Experimental evaluation of EASG}
\label{sec:evaluation}
EASG is evaluated from four perspectives: convergence, results quality, stability, and time scalability. All presented results were obtained in $30$ independent runs per game instance, with parameters and implementation setup discussed in the previous section. In total EASG assessment was based on $9\;000$ trials ($150$ WHG, $90$ SEG and $60$ FIG, each tested $30$ times). All tests were run on Intel Xeon Silver 4116 @ 2.10GHz with 256GB\ RAM.

\subsection{Convergence}
\label{sec:convergence}
\begin{figure}[t]
\centering
  \begin{minipage}[b]{.35\columnwidth}
    \centering
    \includegraphics[width=\linewidth]{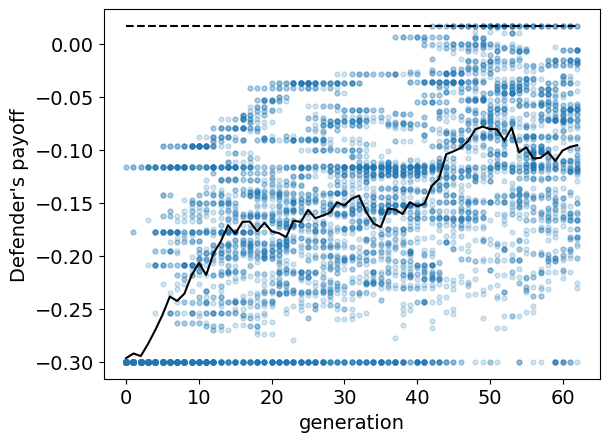}
  \end{minipage}
  \begin{minipage}[b]{.35\columnwidth}
    \centering
    \includegraphics[width=\linewidth]{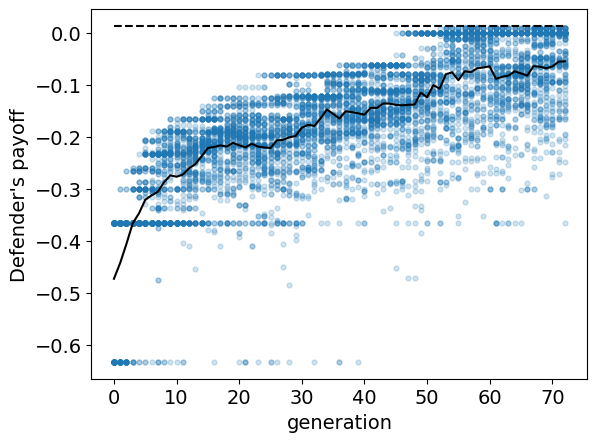}
  \end{minipage}
  \caption{Visualization of typical EASG performance characteristics. The figures present payoffs of all chromosomes in consecutive generations for two sample games. Dashed lines represent the optimal solutions (Defender's payoffs in SStE), solid curves denote the average payoffs across all individuals (in a given generation).}%
  \label{fig:convergence}
  \end{figure}

Figure~\ref{fig:convergence} visualizes two typical convergence characteristics of EASG. The payoffs of random strategies in the initial population are usually centered around one (left figure) or two (right figure) values. In subsequent generations the payoffs are more scattered but, generally, the mean Defender's payoff increases in time, which means that the entire population moves towards the areas with higher payoffs. At the same time, some low-payoff individuals exist in practically all generations due to the use of mutation operator which leads to exploration of new strategies.


%

In none of the experiments EASG attained the maximum number of generations $n_g$ (set to $1\;000$) and in all runs terminated because of the other stopping condition - no improvement of the best-found solution in consecutive $n_c$ generations (set to $20$).

\begin{figure}
\centering
  \subfloat[]{
    \includegraphics[width=.35\columnwidth]{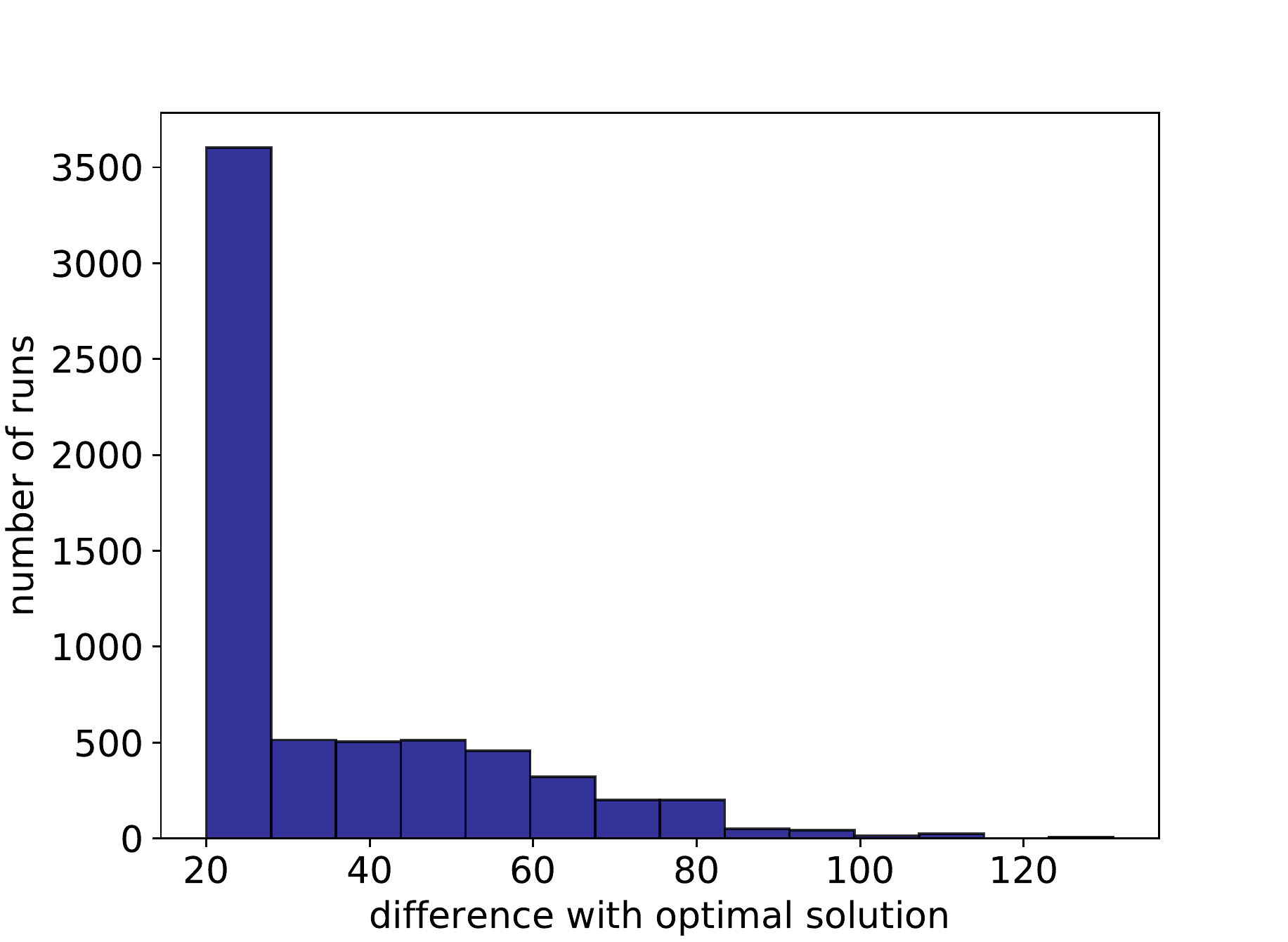}%
    \label{fig:histA}
  }
  \subfloat[]{
    \includegraphics[width=.35\columnwidth]{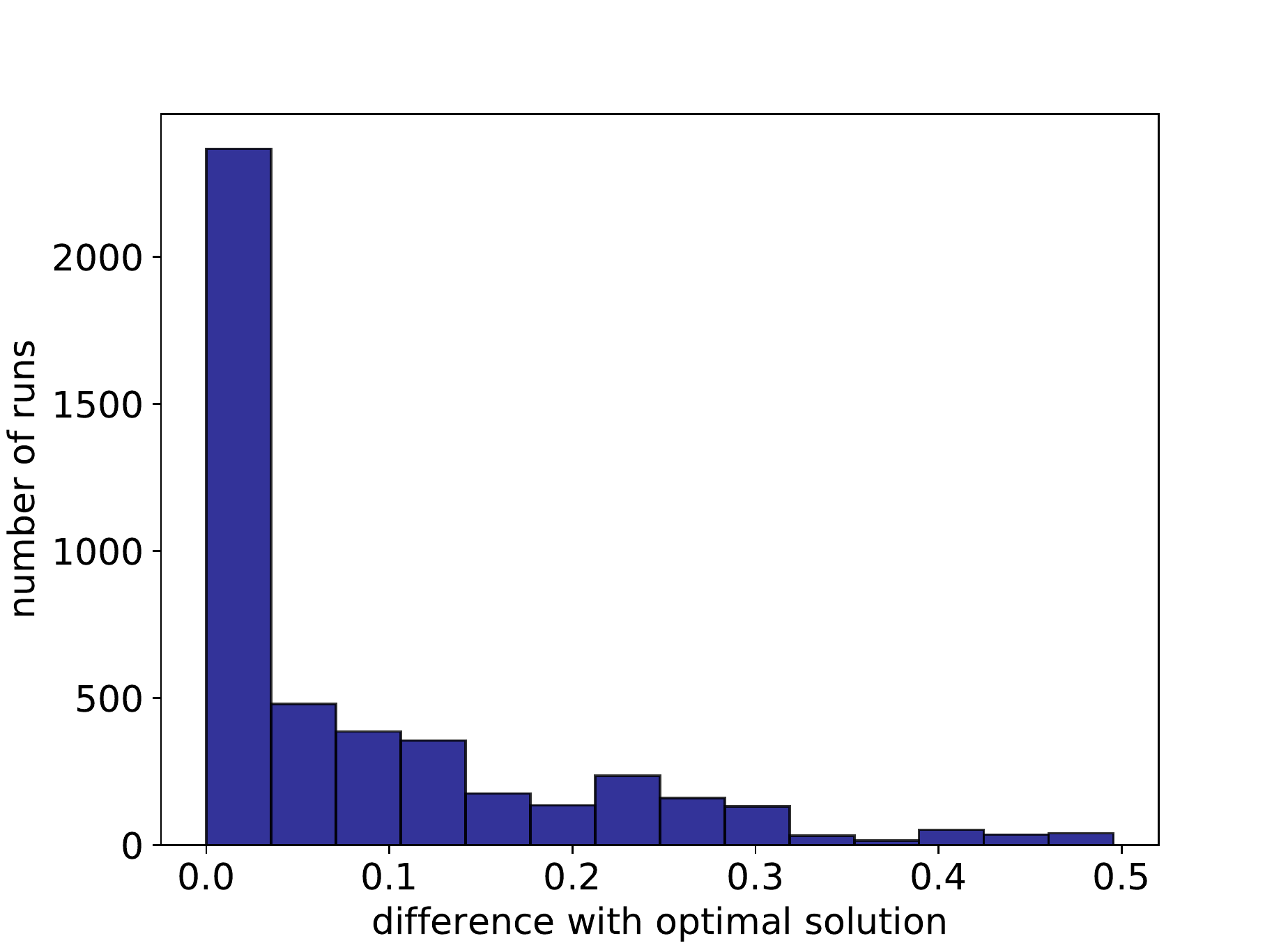}%
    \label{fig:histB}
  }
  \caption{(a): Histogram of the number of generations (for all games) before EASG termination. (b): Histogram of differences between the optimal Defender's payoffs and the payoffs obtained by EASG (for all games with known optimal solutions).}%
  \label{fig:histograms}
  \end{figure}

Figure~\ref{fig:histA} presents a histogram of the numbers of generations in all experiments. The maximum value of $131$ was achieved in one of the $8$ step WHG instances. In more than half of the cases, the number of generations was less than $30$, which means that the final solution was found within the first $10$ generations. This observation supports the claim about generally fast and stable convergence of EASG.
%

\subsection{Results quality}
\label{sec:payoffs}
EASG as an approximate algorithm, obviously, does not guarantee finding optimal solutions. However, experimental evaluation shows that in the majority of the cases the method yields optimal strategies (SStE ones) and in the rest of them, a distance to the optimal solution is narrow.

In order to calculate the optimal solutions for considered game instances, both exact methods (BC2015 and C2016) were run with the limit of 200h per game. If none of the methods were capable of finding the SStE strategy for the Defender within allotted time, the game was excluded from the evaluation of results quality. In effect, $100$ WHG (out of $150$), $60$ SEG (out of $90$) and $45$ FIG (out of $60$) instances were used in the EASG quality assessment.

A histogram of the differences between the optimal solution and the one provided by EASG in all runs, across all tested game instances with known optimal solutions is presented in Fig.~\ref{fig:histB}.

Please note that FIG final payoffs are summed over all controlled vertices, in all rounds. Hence, in order to assure a direct comparability with the results achieved for other game types, they were normalized to $[-1;1]$, i.e. the lowest and the highest possible payoffs (computed by the exact method by minimizing and maximizing the expected Defender's payoff, resp.) were mapped to $-1$ and $1$, respectively.
%

\subsubsection{WHG}
In the case of WHG, both exact methods were able to calculate the SStE for $100$ games with $3-6$ time steps. In all tests involving larger games (with $7$ or $8$ steps) the solution could not be reached due to extensive time
requirements. For $72$ out of the above $100$ games EASG obtained optimal solutions. The mean difference between the optimal and EASG results was equal to $0.0013$ and the highest difference equaled $0.0127$ ($3.7\%$ of the possible payoff range).

\subsubsection{SEG}
For SEG, optimal solutions are known for $60$ games with $4$ and $5$ steps out of which EASG found optimal strategies in $28$ cases ($47\%$). The average divergence from the optimal results equaled $0.0253$ with the highest value of $0.0955$ ($12.2\%$ of the possible payoff range).

In spite of satisfying results, it should be noted that they are noticeably worse than those for WHG. The reason for that is most probably much larger strategy representation - SEG require strategies for several (more than one) Defender's units, additionally with specific variants for the cases of discovering traces, so the effective strategy search space is much wider than that of WHG.

\begin{figure}[h]
	\begin{center}
    \includegraphics[width=0.35\columnwidth]{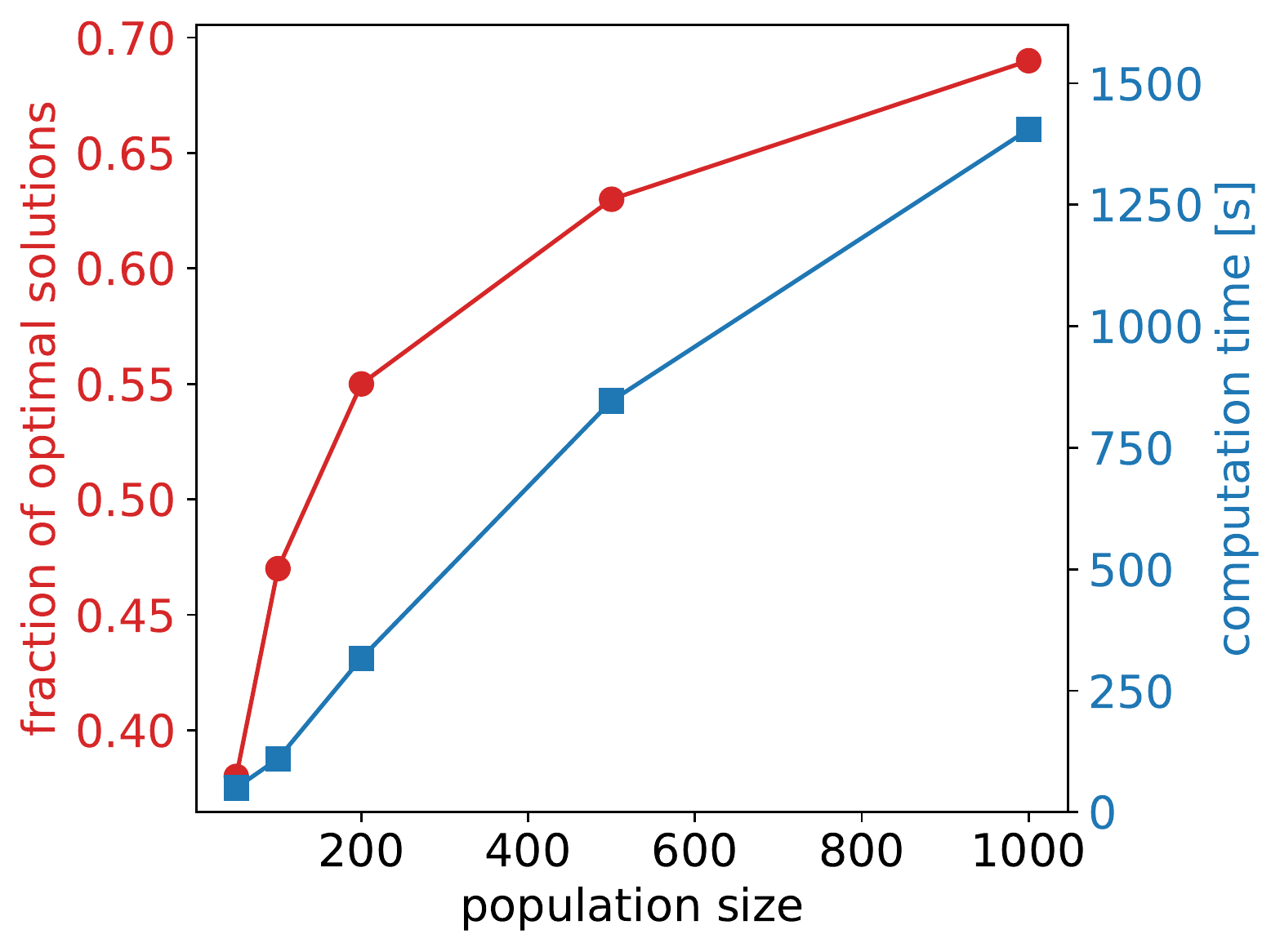}
    \caption{A fraction of SEG instances for which EASG found optimal solutions (circles) and computation time requirements (squares) with respect to the population size.}
    \label{fig:SG_pop_size}
	\end{center}
\end{figure}

In order to verify this hypothesis, additional experiments with bigger populations were executed. If the reason for a relative deterioration of results is indeed SEG's larger strategy space and reacher solution representation, increasing population size should cause a more extensive search of the strategy space and yield better results. The outcomes of these experiments, presented in Figure~\ref{fig:SG_pop_size}, confirm that with greater populations EASG was able to find optimal solutions for a higher percentage of SEG instances, albeit at the cost of computation time increase.

\subsubsection{FIG}
The third set of games - FIG are also recognized as a challenging task for SGs solution methods due to their extensive search space. Please note that FIG game trees grow very fast even for small graphs since in each round (except for the first one in which some restrictions on the Attacker's choice apply) any of the two players can attempt to flip any of graph vertices (not only a successor of his/her current position). For a graph with $n_v$ vertices, in each non-initial round, each of the players has $n_v$ available actions, so the number of nodes in a game tree exceeds $n_v^{2n-1}$, where $n$ is the number of rounds.

Despite large strategy space, EASG managed to achieve optimal solutions in $73\%$ of the cases (exact methods were capable of finding solutions for $45$ test games, out of which EASG yielded the same solutions for $33$ games). The average divergence from the optimal results (normalized in a way described above) equaled $0.0087$ with the highest value of $0.0321$ ($6.4\%$ of the possible payoff range).

The quality of results for FIG is comparable to that for WHG and higher than in the case of SEG. Again, the reason for this results disparity is attributed to a much complex chromosome representation in SEG. While in FIG and WHG solutions (leader's mixed strategies) are encoded straightforwardly, in SEG each chromosome contains separate strategies for each unit and each possible trace discovery scenario (depending on both the vertex and time step). Hence, finding optimal strategy in such a complex space representation is more difficult in a stochastic strategy changing process performed by EASG operators.
\begin{figure*}[ht]
\centering
  \subfloat[Warehouse Games.]{
    \includegraphics[width=.3\columnwidth]{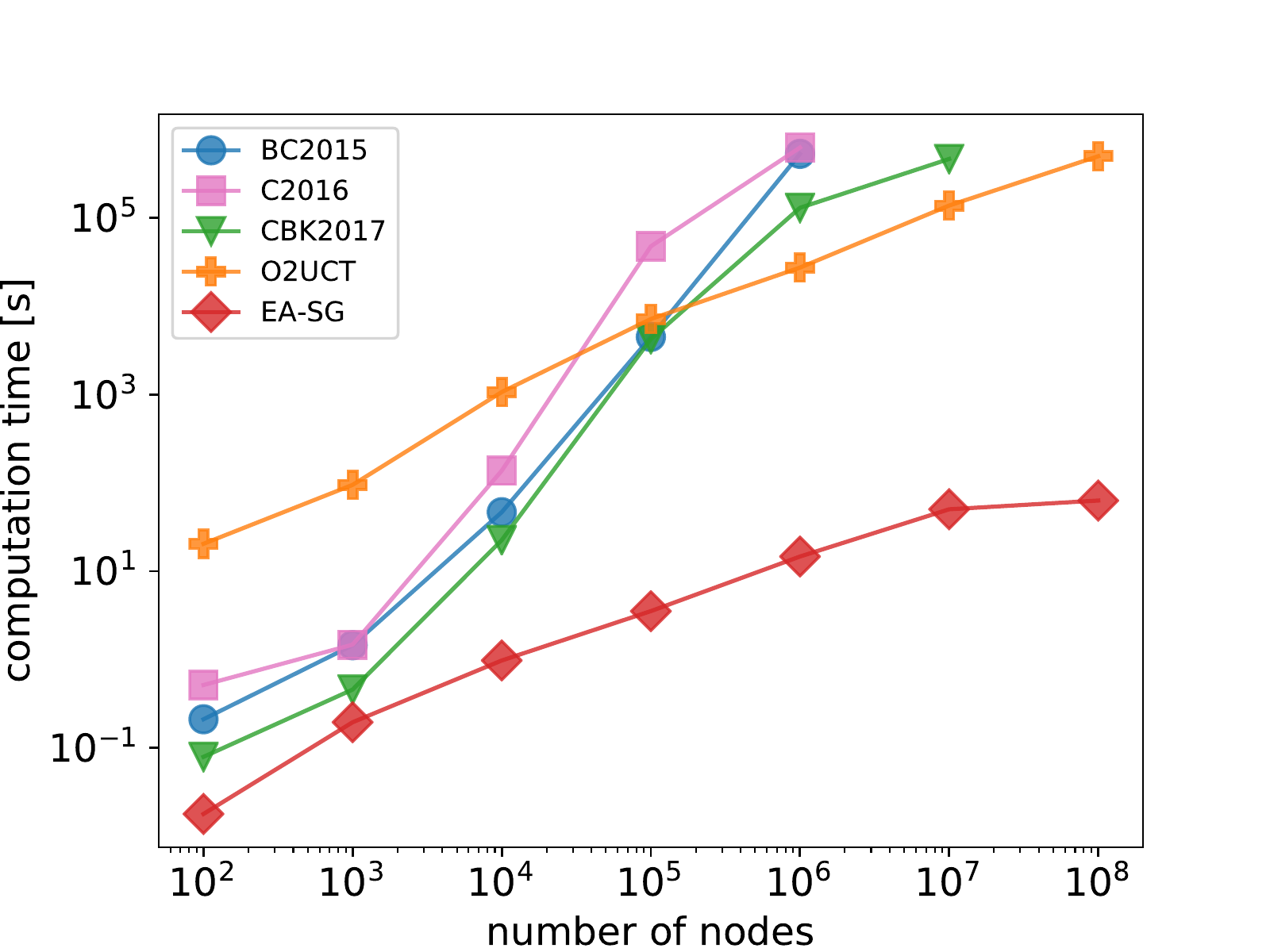}%
    \label{fig:time_wh}
  }
  \quad
  \subfloat[Search Games.]{
    \includegraphics[width=.3\columnwidth]{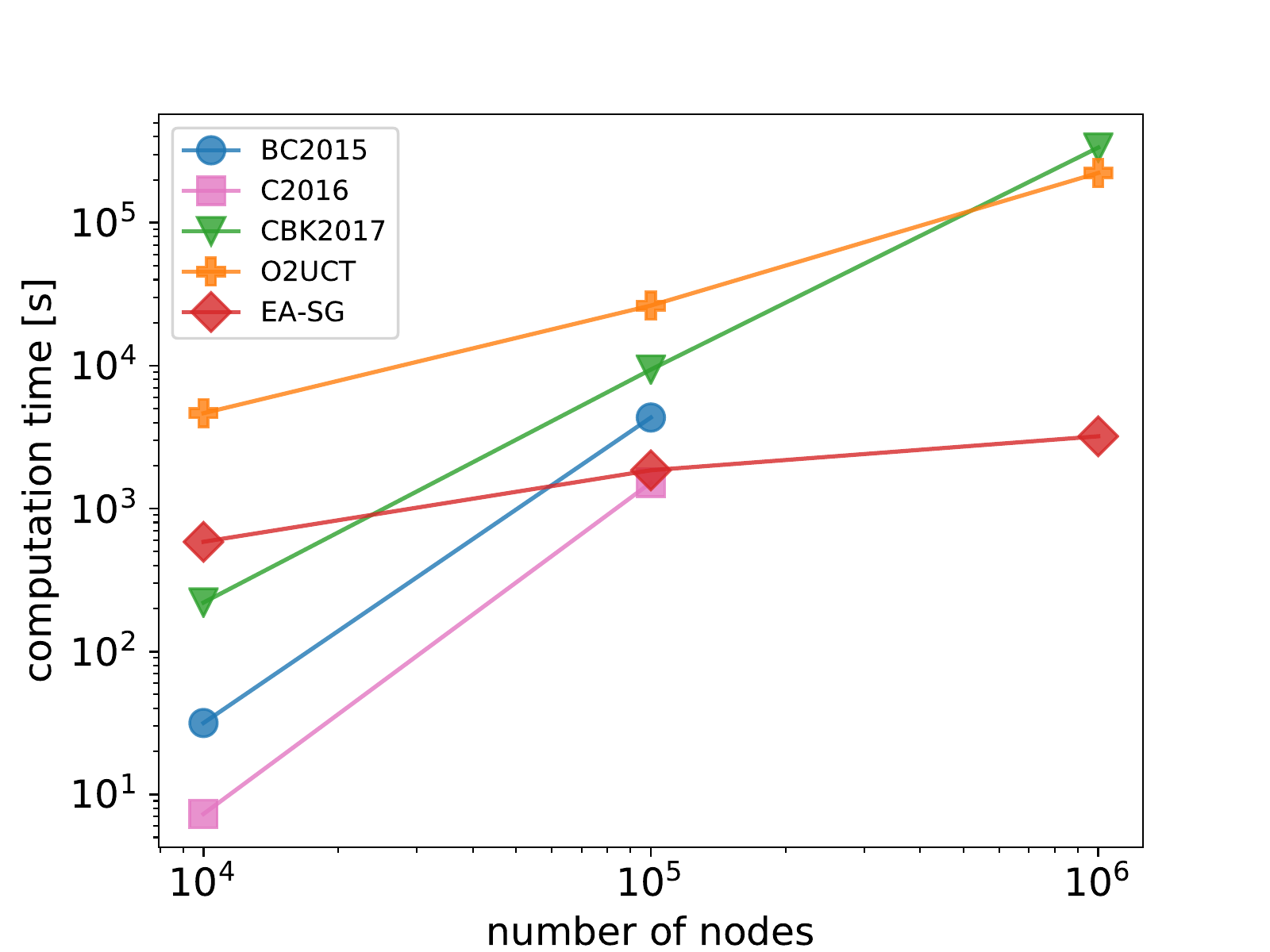}%
    \label{fig:time_sg}
  }
  \quad
  \subfloat[FlipIt Games.]{
    \includegraphics[width=.3\columnwidth]{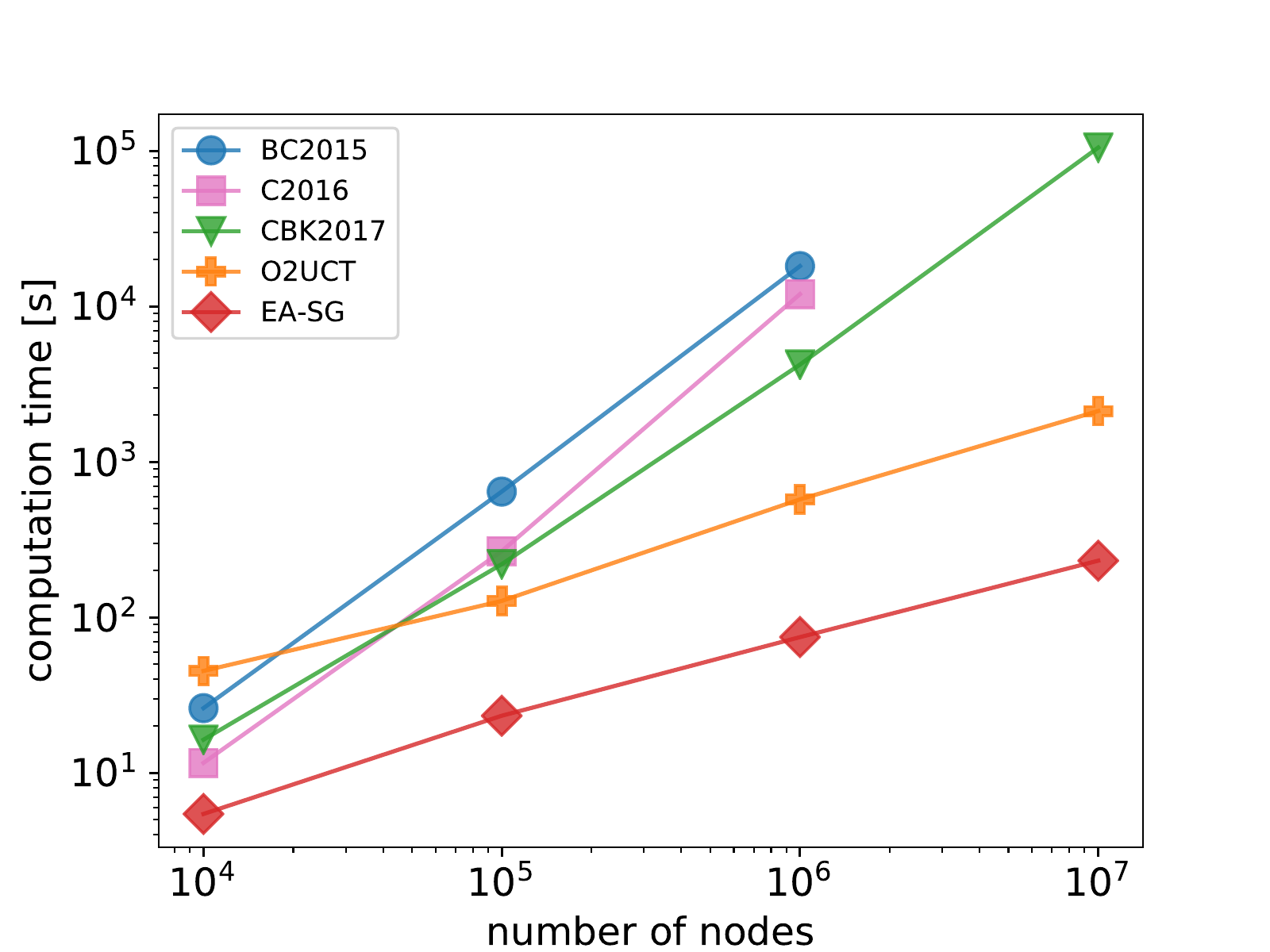}%
    \label{fig:time_fg}
  }
  \caption{Comparison of EASG time scalability vs state-of-the-art methods. Please note a logarithmic scale on both axes.}%
  \label{fig:time_EASG}
\end{figure*}

\subsection{Stability}
\label{sec:stability}
The EASG approach, typically for EA methods, is highly non-deterministic (the initial population, both operators, as well as the selection procedure contain random factors). Hence, the sole ability to obtain optimal solutions, discussed in the previous subsection, is not sufficient for the comprehensive EASG assessment. An equally important aspect of the algorithm is its ability to reproduce good results.

In order to check stability of EASG, standard deviations were computed for all games (including those omitted in the previous subsection due to the lack of exact solutions) across $30$ runs for each game. For $45\%$ of games, standard deviation was equal to $0$ which means perfect stability. The mean standard deviation equaled $0.0059$ with the maximal value $0.1629$ ($36.7\%$ of the possible payoff range).

Among $133$ games for which the optimal solutions were found, the best solution was repeated in all $30$ runs in $82$ ($62\%$) cases. For $96$ of these games ($72\%$) the optimal solution was found in more than $90\%$ of runs. At the other extreme, in $23$ cases ($17\%$) the optimal strategy was achieve only once. The above figures indicate that, despite certain level of randomness, the method offers relatively high stability and is capable of regularly reproducing the optimal solution in the vast majority of the tested games.
%
%
%
%
%

\subsection{Time scalability}
\label{sec:time}
Parameter tuning presented in Section~\ref{sec:parameters_tuning} indicates that time performance of EASG strongly depends on selected steering parameters. Hence, their adjustment allows establishing the expected balance between computation time and quality of results.

Figure~\ref{fig:time_EASG} compares time efficiency of EASG (with parameters setting listed in Table~\ref{tab:parameters}) vs four state-of-the-art algorithms summarized in Section~\ref{sec:state-of-the-art}. First, the games of a given type (separately WHG, SEG and FIG) were divided into subsets of instances with pairwise comparable game tree sizes (after rounding pairwise equal to the nearest power of $10$). Then, for each subset the running times of all game instances belonging to that subset were averaged and plotted. As we mentioned in section~\ref{sec:payoffs}, due to exceeding time limit of $200$ hours per trial, for biggest games the results of exact methods (BC2015 and C2016) could not be plotted.


Computation times of all five methods grow exponentially with respect to the game length. However, in the case of MILP-based algorithms (BC2015, C2016 and CBK2018) the exponents are higher than in the case of approximate approaches (O2UCT and EASG). Overall, EASG demonstrates the highest time efficiency among the tested methods.


\section{Conclusions}
\label{sec:summary}

The paper presents a novel Evolutionary Algorithm approach (named EASG) to solving sequential Security Games. The method explores the space of Defender's strategies by means of evolving a population of candidate strategies in order to find Strong Stackelberg Equilibrium. Proposed approach is generic and can be easily adapted to various types of SGs. Experimental evaluation, performed on games of $3$ different types, with $300$ instances in total, proved robustness and time efficiency of EASG. In more than half of the tested game instances optimal solutions were found and for the rest of them the distance between strategies found by EASG and the optimal ones was very low.

Comprehensive experiments demonstrate that EASG scales in time visibly better than state-of-the-art approaches. Consequently, the proposed method can be applied to bigger games (in terms of the size of a game tree or Defender's strategy space) than state-of-the-art MILP methods. Even though the convergence properties of EASG cannot be mathematically guaranteed (as is the case of the vast majority of EA-based approaches), promising experimental results make us to believe that the method presents a viable alternative to state-of-the-art algorithms when solving larger and more complex sequential SGs.

Another property of the proposed method, which is of a special value in time-critical applications,  is its iteration-based construction. Execution of EASG can be interrupted at any moment and still a valid solution (the best one found so far) will be returned, which makes the approach essentially an \emph{anytime} method.

Most of the EASG computation time is devoted to evaluation of chromosomes due to the necessity of calculation of their payoff against all possible Attacker's pure strategies. Although it is not a serious issue in many real-life applications (where the Attacker's strategy space is usually of moderate size), in some cases (e.g. for games with continuous Attacker's strategy space) it may cause certain inefficiency of the method. Therefore, our current work is concentrated on mitigating this potential problem by
designing a co-evolutionary approach in which a population of Defender's mixed strategies would compete with a population of the Attacker's strategies, so as to provide the optimal Stackelberg Equilibrium strategies for both players.

\section*{Acknowledgments}
  This work was supported by the National Science Centre, grant number 2017/25/B/ST6/02061.

\bibliographystyle{unsrt}  
\bibliography{EA-SG}  

\end{document}